\documentclass[journal]{IEEEtran}
\usepackage{cite}
\usepackage{graphicx}
\usepackage{amsmath,amssymb,amsfonts}
\usepackage{algorithm, algorithmic}
\usepackage{subfigure}
\usepackage{multirow}
\usepackage{array}
\usepackage{color}
\usepackage{verbatim}
\usepackage{CJKutf8}
\usepackage[normalem]{ulem}

\usepackage{pgfplots}
\usepackage{pgfplotstable}
\pgfplotsset{compat=1.17}

\newif\ifrevision
\newcommand{\highlight}[1]{%
\ifrevision
\textcolor{blue}{#1}%
\else
#1%
\fi
}

\begin{document}

\revisionfalse

\title{Deep Joint Source-Channel Coding\\ for Semantic Communications}
%
%
%

\author{Jialong Xu,
		Tze-Yang Tung,
        Bo Ai,
        Wei Chen,
        Yuxuan Sun
        and~Deniz G{\"u}nd{\"u}z
    \thanks{For the purpose of open access, the authors have applied a Creative Commons Attribution (CCBY) license to any Author Accepted Manuscript version arising from this submission.}
\thanks{This work is supported by the Natural Science Foundation of China (62122012, 62221001); the Beijing Natural Science Foundation (L202019, L211012, L222044), the Fundamental Research Funds for the Central Universities (2022JBQY004, 2022JBXT001), the Talent Fund of Beijing Jiaotong University (2023XKRC030), and from the UKRI through projects AIR (ERC-CoG, EP/X030806/1) and SONATA (EPSRC-EP/W035960/1).

\textit{(corresponding authors: Bo Ai; Wei Chen)}}
\thanks{Jialong Xu is with State Key Laboratory of Advanced Rail Autonomous Operation, Beijing Jiaotong University, China and also with Frontiers Science Center for Smart High-speed Railway System. (e-mail: jialongxu@bjtu.edu.cn)}
\thanks{Bo Ai is with State Key Laboratory of Advanced Rail Autonomous Operation, Beijing Jiaotong University, China, Beijing Engineering Research Center of High-speed Railway Broadband Mobile Communications, and also with School of Information Engineering, Zhengzhou University, Zhengzhou, China. (e-mail: boai@bjtu.edu.cn)}
\thanks{Tze-Yang Tung and Deniz G{\"u}nd{\"u}z are with the Department of Electrical and Electronic Engineering, Imperial College London, London SW7 2BT, UK (e-mail: \{tt2114, d.gunduz\}@imperial.ac.uk).}
\thanks{Wei Chen is with State Key Laboratory of Advanced Rail Autonomous Operation, Beijing Jiaotong University, China and also with Key Laboratory of Railway Industry of Broadband Mobile Information Communications. (e-mail: weich@bjtu.edu.cn)}
\thanks{Yuxuan Sun is with the School of Electronic and Information Engineering, and the Frontiers Science Center for Smart High-Speed Railway System, Beijing Jiaotong University, Beijing, China. (e-mail: yxsun@bjtu.edu.cn)}
}

\markboth{Journal of \LaTeX\ Class Files,~Vol.~X, No.~Y, November~2022}%
{Xu \MakeLowercase{\textit{et al.}}: Deep Joint Source-Channel Coding for Semantic Communications}

\maketitle

\begin{abstract}
\highlight{Semantic communications is considered as a promising technology to increase the efficiency of next-generation communication systems, particularly targeting human-machine and machine-type communications.} In contrast to the source-agnostic approach of conventional wireless communication systems, semantic communication seeks to ensure that only the relevant information for the underlying task is communicated to the receiver. Considering that most semantic communication applications have strict latency, bandwidth, and power constraints, a prominent approach is to model them as a joint source-channel coding (JSCC) problem. Although JSCC has been a long-standing open problem in communication and coding theory, remarkable performance gains have been shown recently over existing separate source and channel coding systems, particularly in low-latency and low-power scenarios. Recent progress is thanks to the adoption of deep learning techniques for \highlight{joint source-channel code design} that outperform the concatenation of state-of-the-art compression and channel coding schemes, which are results of decades-long research efforts. In this article, we present an adaptive deep learning based JSCC (DeepJSCC) architecture for semantic communications, introduce its design principles, highlight its benefits, and outline future research challenges that lie ahead.

\end{abstract}

\begin{IEEEkeywords}
Joint source-channel coding, semantic communications, deep learning.
\end{IEEEkeywords}

%
\IEEEpeerreviewmaketitle

\section{Introduction}
\label{Introduction} 
\IEEEPARstart{A}{lmost} all existing communication systems, including the past five generations of mobile communication standards, follow the layered design approach, where sampling, quantization and compression are taken care of by the application layer, while the physical layer takes care of the communication aspects. In this design paradigm, physical layer coding and modulation are designed independently of the application they serve; their goal is to convey packets of bits to their respective receivers as reliably as possible within the prescribed latency, bandwidth, and power limitations, without taking into account how these bits are used at the receiver end. This paradigm has its roots in Shannon's well-known Separation Theorem \cite{shannon1949mathematical}, which proves that separate source and channel coding is without loss of optimality in the usual Shannon theoretic asymptotic regime of infinite block length and unbounded complexity. 


\begin{figure}[t]
\centering
\includegraphics[width=1\columnwidth]{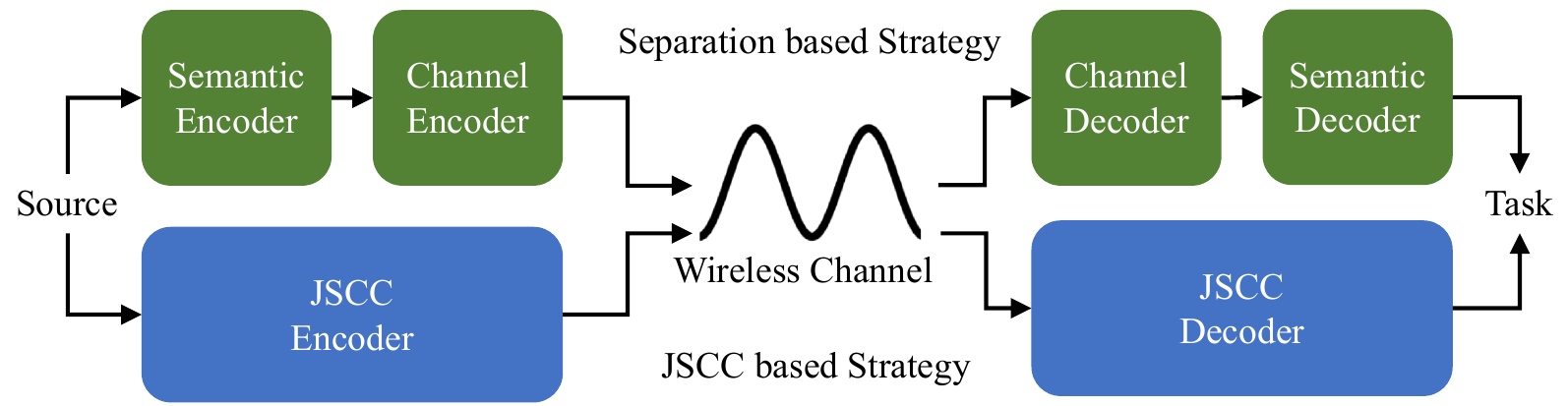}
\caption{Separation based strategy vs. JSCC for semantic communications.}
\label{Fig:arch}
\end{figure}

Separation-based designs have been satisfactory for our current communication networks whose main purpose is content delivery for human consumption. However, a significant transformation in the data traffic is expected in the next generation of mobile networks. \highlight{Emerging applications, such as industrial robotics, autonomous driving, drone networks, virtual reality and metaverse, all have significantly different communication requirements and constraints.} First of all, signals are delivered to other machines for inference purposes, rather than for reconstruction. This means that significant reduction in traffic load is possible by taking the signal semantics into account. Additionally, these applications impose stringent constraints on latency, computational capability, bandwidth usage, and energy consumption. It is becoming increasingly evident that the desired performance requirements cannot be met within the boundaries of the conventional separate network architecture. Although the fifth generation (5G) of mobile networks has defined the ultra reliable low latency communications (URLLC) scenario to deal with the latency problem, it often ignores the complexity of signal processing (e.g., compression) that can easily dominate the end-to-end latency. 

An alternative design paradigm has started to emerge in recent years (see Fig. \ref{Fig:arch}). In joint source-channel coding (JSCC), the transmitter directly maps the source signal to channel symbols, and the receiver recovers its estimate directly from the noisy channel output. In the JSCC paradigm, bits are no longer the common currency between the application and physical layers. Although JSCC is known to outperform separate source and channel coding in the finite block length regime, a practical JSCC scheme with reasonable complexity and desirable performance has remained elusive due to the complexity of designing such a scheme. 
Most existing JSCC solutions combine conventional source and channel code designs, and jointly optimize their parameters for improved end-to-end performance. On the other hand, what we really need is a transformation from the source signal space to the channel input space (and vice versa at the receiver), where similar input signals are mapped to similar channel inputs so that the receiver can recover a reconstruction of the source signal with minimal distortion despite the noise and other impairments of the channel. While this is a highly challenging optimization problem to solve in a model-driven manner, it was shown in \cite{bourtsoulatze2019deep} that neural networks can be trained to successfully learn such a transform directly from data. In the so-called DeepJSCC design paradigm introduced in \cite{bourtsoulatze2019deep}, the encoder and decoder are parameterized as deep neural networks (DNNs), forming an \textit{autoencoder} with channel noise injected into the latent space, and optimized jointly in a data-driven manner.



The end-to-end training approach of DeepJSCC lends itself well to the notion of \emph{semantic communications} that has been receiving significant research attention from both academia and industry in recent years \cite{JSAC:semantics}. Here, we refer to semantic communication as conveying the most relevant information prescribed by the underlying task. For example, if the goal is to send text while maintaining its semantics, the particular language and communication context will specify the fidelity function, i.e., which reconstructions are accepted as semantically equivalent for each input sentence, and the goal will be to communicate while maintaining within those semantic constraints. Similarly, in transmitting an image, the application will specify if the receiver wants a reconstruction that has high fidelity at the pixel level, or more interested in reconstructing a perceptually natural looking image, or only interested in reconstruction of the classes of the objects in the image. In each case, the application specifies the distortion/fidelity function. In the DeepJSCC approach, since the encoder/decoder pair can be trained with any desired loss function, they can learn not only to extract relevant semantic features for the end-to-end communication task, but also to communicate them over the noisy channel with as high fidelity as possible. Yet, it is worth noting that semantic communication \emph{per se} does not require JSCC, and a separation-based system can also benefit from semantic compression. On the other hand, as we will show through examples in this paper, significant gains can be achieved by a joint design. 

An important advantage of DeepJSCC is the graceful degradation with respect to channel quality. Conventional digital systems suffer from the \emph{cliff-effect}: communication completely breaks down if the channel quality falls below the correction capability of the channel code. Although these errors are often compensated by methods such as hybrid automatic repeat request (HARQ), they consume additional bandwidth and result in further delay. In contrast, DeepJSCC not only improves the end-to-end performance for a target channel quality, but also exhibits \textit{graceful degradation} as the channel quality decays. This property of DeepJSCC has important implications in terms of channel estimation requirements when communicating over a time-varying channel.

The rest of the paper is organized as follows. In Section \ref{sec:adaptive_architecture}, we present an adaptive DeepJSCC architecture for semantic communication of images, and show its remarkable performance. We then discuss the security implications of DeepJSCC in Section \ref{sec:security}, and offer possible solutions. In Section \ref{sec:other_applications}, we present applications of the DeepJSCC paradigm beyond image transmission.
Finally, we conclude the paper, and highlight important research challenges in Section \ref{sec:conclusions}.

\begin{figure}[t]
\centering
\vspace{-.3in}
\includegraphics[width=1\columnwidth]{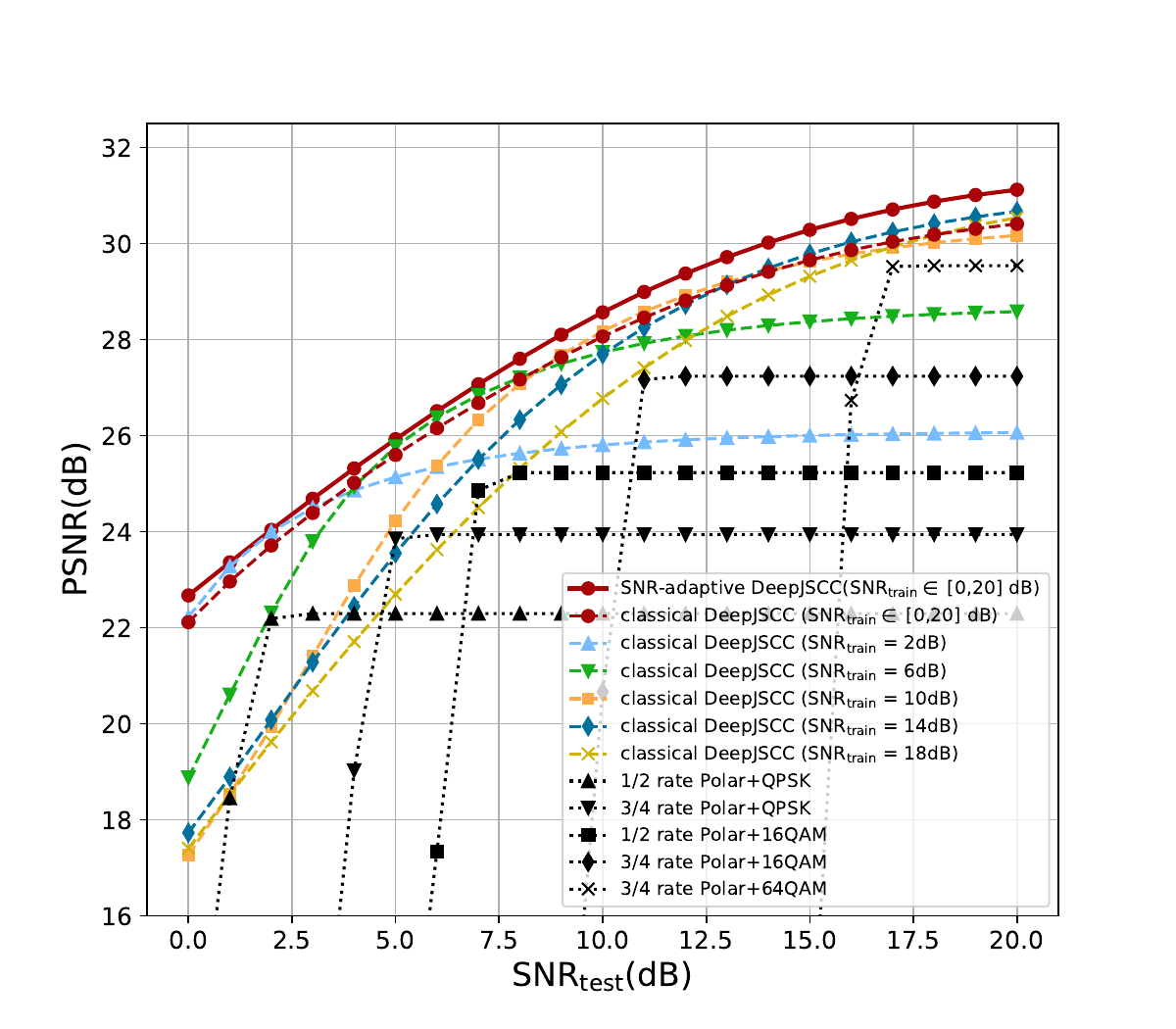}
\caption{Comparison of SNR-adaptive DeepJSCC and classical DeepJSCC on the CIFAR-10 dataset.}
\label{Fig:c8_awgn_cifar10}
\end{figure}

\section{An Adaptive DeepJSCC Architecture}
\label{sec:adaptive_architecture}

Recent works have demonstrated the superiority of DeepJSCC over separate source and channel coding for various information sources, e.g., image \cite{bourtsoulatze2019deep}, video \cite{tung2022deepwive}, text \cite{xie2020lite} and speech \cite{han2022semantic}. We will focus on image transmission in this paper to illustrate the benefits of DeepJSCC for various tasks.

\begin{figure*}[t]
\centering
\includegraphics[width=1.6\columnwidth]{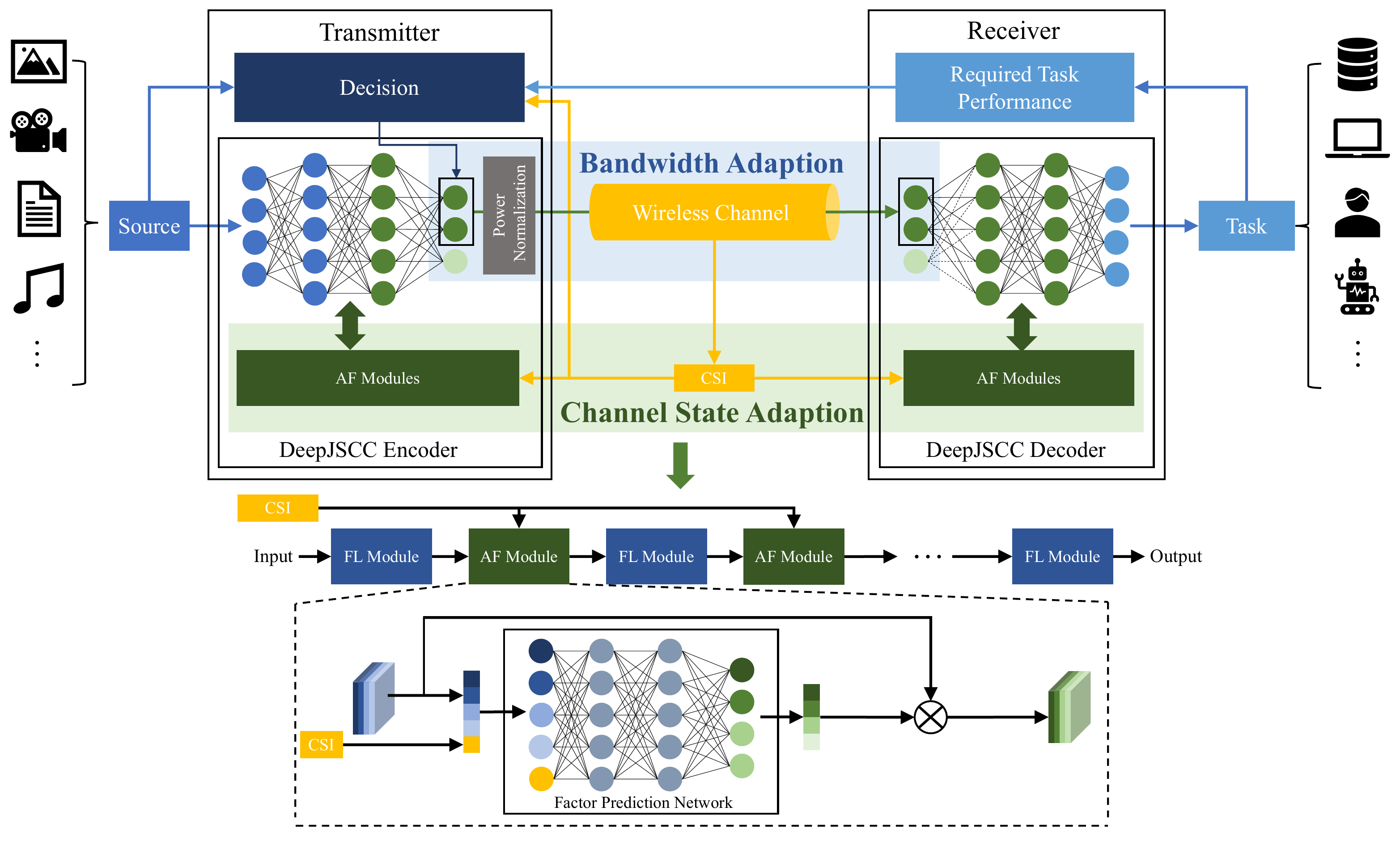}
\caption{The adaptive DeepJSCC architecture for end-to-end semantic communications.}
\label{Fig:adjscc}
\end{figure*}

In the original design in \cite{bourtsoulatze2019deep}, DeepJSCC encoder employs a fully convolutional neural network (CNN) architecture to transform the input image $\boldsymbol{x} \in \mathbb{R}^n$ to a complex channel codeword $\boldsymbol{z} \in \mathbb{C}^k$, under an average power constraint. The receiver employs a matching CNN to transform the received noisy vector $\boldsymbol{\hat{z}} \in \mathbb{C}^k$ to reconstruction $\boldsymbol{\hat{x}} \in \mathbb{R}^n$. We call the dimension of the source signal $n$ as the \textit{source bandwidth} and that of the output signal $k$ as the \textit{channel bandwidth}. \textit{Bandwidth ratio} is defined as $\rho \triangleq k/n$. The power of DeepJSCC comes from the fact that it learns a communication scheme from scratch by optimizing all transformations in a data-driven manner, with a non-trainable differentiable channel model in the bottleneck layer. This simplifies the JSCC design procedure, and allows adaptation to any particular source statistics,  channel model, and fidelity measure specified by the end task. 

In Fig.~\ref{Fig:c8_awgn_cifar10}, we present the peak signal-to-noise ratio (PSNR) performance achieved by DeepJSCC after being trained on a portion of the CIFAR-10 dataset, transmitted over an additive white Gaussian noise (AWGN) channel with a channel bandwidth of $k=256$ symbols. Inputs are colored images of size $32 \times 32$, which corresponds to a bandwidth ratio of $\rho = 256/(3 \times 32 \times 32)= 1/12$. Each of the curves in the figure, denoted by `classical DeepJSCC', corresponds to an encoder/decoder pair proposed in \cite{kurka2020deepjscc} trained on a different channel signal-to-noise ratio (SNR), $\mathrm{SNR}_{\mathrm{train}}$, and tested over a range of channel conditions, $\mathrm{SNR}_{\mathrm{test}} \in [0, 20]~\mathrm{dB}$. We can clearly observe that the performance of DeepJSCC gracefully degrades as the test SNR drops below the training SNR, and it slowly improves when it gets better. This is in contrast with the `cliff edge' behaviour of the separation based schemes in the figure, where the performance sharply falls when the channel quality falls below the SNR threshold required to reliably decode the channel code. The separation based schemes in the figure employ BPG codec for compression and polar codes with QAM modulation for transmission. We refer the readers to \cite{bourtsoulatze2019deep} for a comparison of these two approaches in terms of the computational complexity. 

We also observe from Fig.~\ref{Fig:c8_awgn_cifar10} that the best performance at each $\mathrm{SNR}_{\mathrm{test}}$ value is achieved by training the encoder/decoder networks at the same SNR. However, this would imply storing separate DNN parameters for every possible channel condition, which would make DeepJSCC impractical. An alternative approach would be to train a single encoder/decoder pair to be used over a range of channel conditions. We see in Fig.~\ref{Fig:c8_awgn_cifar10} that DeepJSCC trained over the whole range of channel SNRs can achieve reasonable performance, and shows the applicability of DeepJSCC when communicating over a time-varying channel when channel state information (CSI) is not available, but it falls short of the performance that can be achieved when CSI is known.

To mitigate this limitation and to bring DeepJSCC one step closer to practice, an SNR-adaptive DeepJSCC architecture is also presented. Here, assuming CSI information at both the transmitter and receiver, DeepJSCC encoder and decoder are enhanced by an attention feature (AF) module \cite{xu2022wireless}. The decision and the required task performance (RTP) modules are also introduced. The RTP module generates a required performance level according to the specific task, e.g., PSNR or multi-scale structural similarity index measure (MS-SSIM) for image or video reconstruction tasks, or BLEU for text reconstruction tasks, and sends it to the transmitter. The decision module at the transmitter infers how much channel bandwidth, $\widetilde{k}$, should be transmitted to achieve the desired performance level, taking into account the CSI and the specific input signal. 

Note that the channel bandwidth is fixed in the classical DeepJSCC architecture. Similarly to SNR adaptation, to realize bandwidth adaptation, we would need to train and store different DeepJSCC networks for different channel bandwidths. In the test stage, the transmitter would choose the one with the same output size as the inferred channel bandwidth from among a set of trained models according to channel conditions. To overcome this problem, adaptive DeepJSCC adopts the architecture with shared weights as proposed in \cite{kurka2021bandwidth} for channel bandwidth adaptation. 

\subsection{Channel State Adaptation}

In order for DeepJSCC to adapt to channel conditions, the AF module is proposed in \cite{xu2022wireless}, which interacts with the DNNs in the DeepJSCC encoder and decoder. As shown in the lower part of Fig.~\ref{Fig:adjscc}, the feature learning (FL) and AF modules are alternately connected. The output of each FL module is combined with the CSI, and fed into the next AF module. The output of each AF module is then fed to the next FL module. Each FL module corresponds to a layer of the classical DeepJSCC architecture, and is responsible for extracting the semantic features most relevant for the underlying task specified through the end-to-end distortion/fidelity function. Given the CSI and the output of the previous FL module (i.e., the FL features), the AF module first extracts information from the FL features and then fuses it with the CSI to form the context information. Next, the factor prediction network produces an attention mask for the FL features with the context information as input, and scales the FL features based on the attention weights. This architecture is trained over a range of SNR values, and the AF module learns to assign different weights to different features under different channel conditions, so that in poor channel conditions only the most important features are communicated with more robustness against channel noise. 

We observe in Fig.~\ref{Fig:c8_awgn_cifar10} that adaptive DeepJSCC performs better than classical DeepJSCC, especially when the $\rm SNR_{test}$ mismatches the $\rm SNR_{train}$. More interestingly, adaptive DeepJSCC outperforms the classical DeepJSCC even when $\rm SNR_{test} = \rm SNR_{train}$ in the high SNR regime, which shows that training at low SNRs provides additional robustness.

\begin{figure}[t]
\centering
\vspace{-.25in}
\includegraphics[width=1\columnwidth]{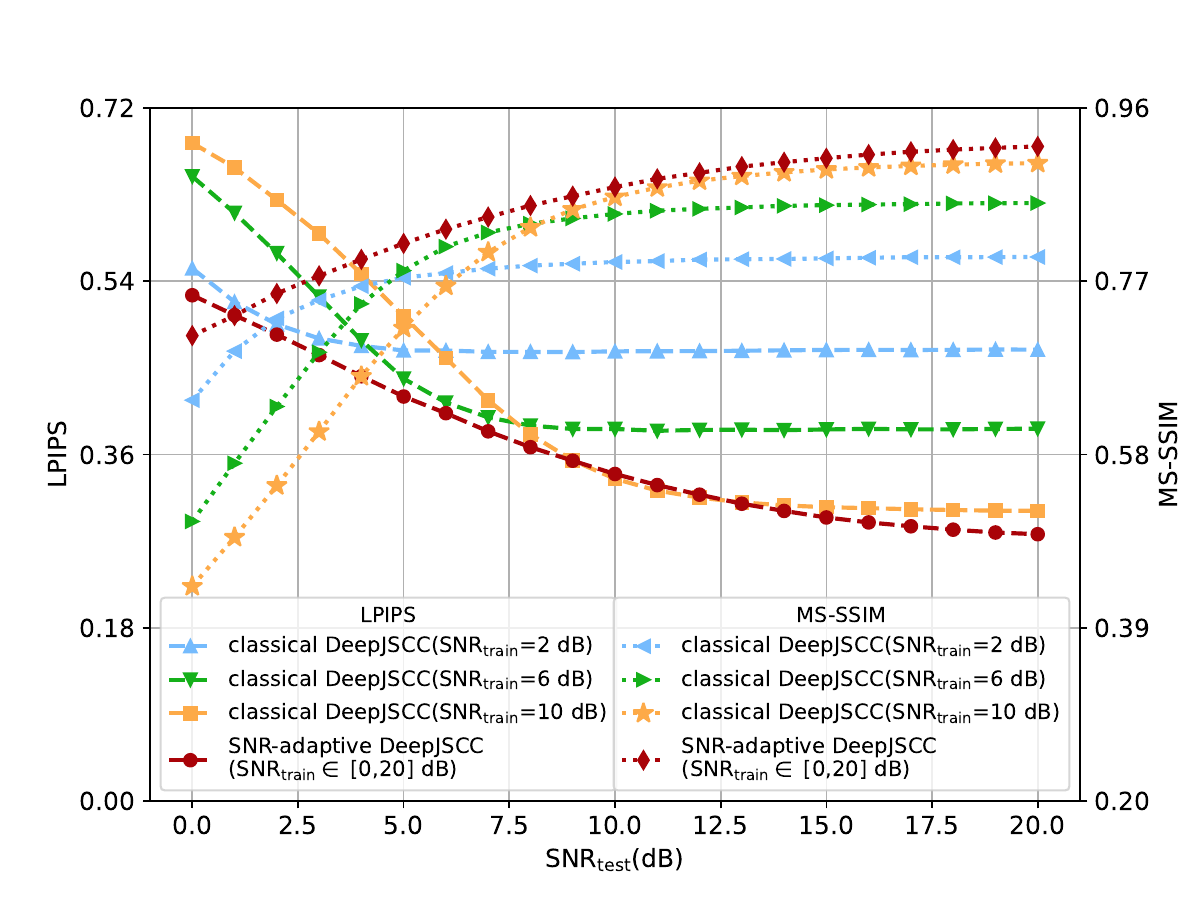}
\caption{\highlight{Comparison of SNR-adaptive DeepJSCC and classical DeepJSCC trained on ImageNet with the semantic loss functions LPIPS and MS-SSIM and tested on the Kodak dataset.}}
\label{fig:deepjscc_Kodak}
\end{figure}

To achieve bandwidth adaptivity, in \cite{kurka2021bandwidth}, the output of the DeepJSCC encoder is divided into multiple layers, and a random number of first $l$ layers are transmitted to the decoder during training. As a consequence, the encoder learns to order the information in the latent vector in descending order of importance relative to the number of layers. This results in a successive refinement scheme, where the more layers are received, the better the reconstructed image.

\highlight{\subsection{DeepJSCC with Semantic Loss}}

\begin{figure*}[t]
\centering
\includegraphics[width=1.8\columnwidth]{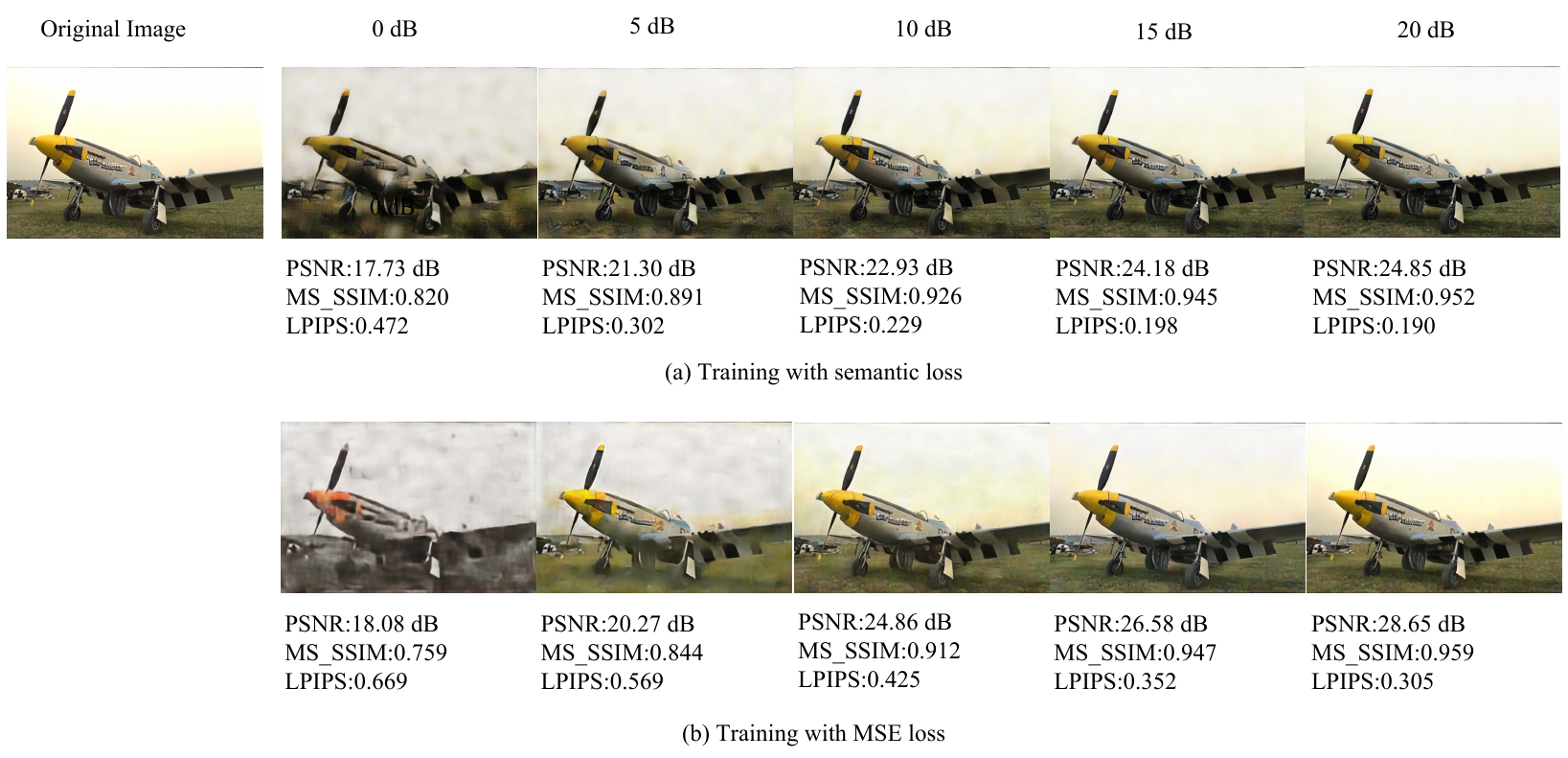}
\caption{\highlight{Visualization of an image from the Kodak dataset transmitted with DeepJSCC over different channel conditions and loss functions.}}
\label{Fig:se_visual}
\end{figure*}

\highlight{While we have considered PSNR performance in Fig.~\ref{Fig:c8_awgn_cifar10}, it is known that large PSNR values do not necessarily mean a high perceptual quality. Many different metrics have been proposed in the literature for better image quality assessment. Thanks to its flexibility, DeepJSCC can be optimized for any semantic loss function. In Fig.~\ref{fig:deepjscc_Kodak}, we consider learned perceptual image patch similarity (LPIPS) and multi-scale structural similarity index measure (MS-SSIM) as loss functions, which are considered to be better aligned with semantic image transmission. We train this model on the ImageNet dataset and evaluate its performance on the Kodak dataset, which consists of relatively large colored images (see Fig. \ref{Fig:se_visual} for an example). A lower LPIPS value or a higher MS-SSIM value represents better perceptional performance. With the assistance of the AF module, SNR-adaptive DeepJSCC trained over an SNR range again outperforms the best LPIPS/MS-SSIM performance obtained by training for a specific SNR value. We also present in Fig.~\ref{Fig:se_visual} the results of the DeepJSCC transmission over different channel qualities trained both for PSNR and LPIPS. We observe that DeepJSCC trained for LPIPS recovers more realistic looking images even though the reconstruction may not be accurate at the pixel level, which can be considered as acquiring better image semantics.}

\section{Security in DeepJSCC}

A critically important issue in DeepJSCC is security against eavesdroppers. In separation-based systems, the source signal is first compressed, then encrypted, and the encrypted data is encoded by the channel encoder. Wellknown encryption methods, e.g., data encryption standard (DES), advanced encryption standard (AES) or Rivest-Shamir-Adleman (RSA) can be employed. The independence of channel coding from the source information allows the adoption of such encryption techniques without effecting the end performance. However, these encryption schemes are not compatible with DeepJSCC, which directly maps the source signal to channel input symbols. Indeed, DeepJSCC benefits from the correlation between the signals transmitted over the channel and the underlying source signal. The channel noise may reduce this correlation, and hence, impair the reconstruction quality, but unlike in digital communication, the system never completely breaks down. This advantage, however, becomes a challenge from a security point of view. Next, we propose two potential solutions for securing DeepJSCC against potential eavesdroppers.  

\subsection{Security of the source signal}\label{sec:security}

The first approach aims at protecting the source signal before it is fed to the DeepJSCC encoder. Conventional source protection methods, e.g., permutation and scrambling, destroy the inherent correlations within the source, causing a performance degradation in subsequent DeepJSCC transmission. In \cite{xu2021deep}, a joint source protection and source-channel coding method, called \emph{DeepJESCC}, is proposed to overcome this problem, where another module is introduced, which transforms the source signal to a protected domain before transmission. 
At the receiver, once the protected source is recovered, an inverse network is employed to recover the source. In order to train both the protection and the transmission networks, a novel loss function consisting of the reconstruction loss, the protection loss, and the inversion loss is proposed \cite{xu2021deep}. Moreover, a feature extraction network is designed for calculating the protection loss between the original source and its protected version at the transmitter, and the deprotection loss between the source and its protected version recovered at the receiver. In the training stage, DeepJESCC minimizes the reconstruction loss and simultaneously maximizes the protection and deprotection losses to jointly train the protection, DeepJSCC, and deprotection networks. Note that DeepJESCC is a general principle without limiting the architectures used; on the other hand, it will rely on the complexity of these networks, and cannot provide security guarantees unless these networks can be protected against leakage, or updated sufficiently frequently to prevent eavesdropping.

\subsection{Securing the transmitted symbols}

Another strategy for securing the physical layer communication of DeepJSCC schemes is to encrypt the transmitted symbols.
In \cite{tung2022deep}, the authors propose using a known public-key cryptographic scheme based on the learning with error (LWE) problem, in order to encrypt the transmitted symbols.
The encryption scheme embeds the message in a finite lattice, which, after perturbation by random noise, makes recovering the original message computationally infeasible without the secret key. With the secret key, however, the decoder recovers the original message with the random noise perturbation.
The proposed scheme, called \emph{DeepJSCEC}, therefore, treats the noise from the encryption scheme as well as from the channel as a compound noise, and learns the DeepJSCC encoder/decoder pair that is robust against both sources of noise.
\emph{DeepJSCEC} not only retains the beneficial properties of  DeepJSCC, such as graceful degradation of image quality with varying channel quality and lower end-to-end distortion, but also provides security against chosen-plaintext attacks. It is also shown in \cite{tung2022deep} that the proposed encryption method is problem agnostic; that is, it can be applied to other end-to-end JSCC problems without modification. Moreover, since the scheme is based on a public-key encryption scheme, new keys can be generated without retraining the DNN models, which makes the scheme highly practical.

\section{Other Applications of DeepJSCC}\label{sec:other_applications}

In the previous sections, we have shown the superiority of DeepJSCC particularly in wireless image delivery. However, DeepJSCC is a highly flexible framework, and it can be used for a wide variety of source and channel distributions and for different objectives. In this section, we introduce three novel applications of DeepJSCC as examples of its significant potential in a wide variety of communication applications.

\subsection{Wireless Video Delivery (DeepWiVe)}

The application of JSCC to video streaming and delivery is expected to have a significant impact given that video traffic constitutes a large portion of all network traffic. The first DeepJSCC approach for wireless video delivery is studied in \cite{tung2022deepwive}. The proposed DeepWiVe architecture faces unique challenges that do not appear in image transmission. In video coding, it is important to exploit the temporal correlations within the frames to reduce the compression rate. In DeepWiVe, however, residual errors can only be estimated at the transmitter due to the random nature of the reconstructed frames at the receiver. Another challenge is bandwidth allocation among frames. In \cite{tung2022deepwive}, this problem is resolved by employing reinforcement learning to sequentially allocate the available channel bandwidth among frames according to their content, e.g., more bandwidth is allocated to those frames with more temporal variations. The authors show that the proposed DeepWiVe scheme can outperform state-of-the-art H.265 video compression codec followed by LDPC channel codes in terms of MS-SSIM. DeepWiVe provides not only better end-to-end perceptual performance and robustness to channel variations, but also requires significantly lower computational complexity compared to common video compression codecs. These aspects make it particularly appealing for virtual and augmented reality applications, video delivery from drones, or video sharing among autonomous vehicles, which require high-quality, low-latency and flexible delivery.

\subsection{DeepJSCC for CSI feedback}

Accurate CSI feedback is crucial for massive multiple-input multiple-output (MIMO) communications, as it significantly improves the capacity and energy efficiency of the system.
However, CSI feedback can also induce a very large overhead for the system particularly due to the large number of antennas and users being served.
Just like the transmission of images or video, CSI feedback schemes built upon the separation of source and channel coding can easily suffer from the cliff-effect, which leads to a cliff-edge drop off in the throughput of the system.
As such, in \cite{mashhadi2019cnnbased, xu2022deep}, DeepJSCC is proposed for CSI feedback.
The idea is to directly map the CSI estimate at the receiver to complex channel symbols of the feedback channel, just like in image transmission. This can be considered yet another case of semantic communications, as the goal is not necessarily to reconstruct the CSI matrix accurately, but to achieve the highest communication rate.



\begin{figure}[t]
\centering
\includegraphics[width=1\columnwidth]{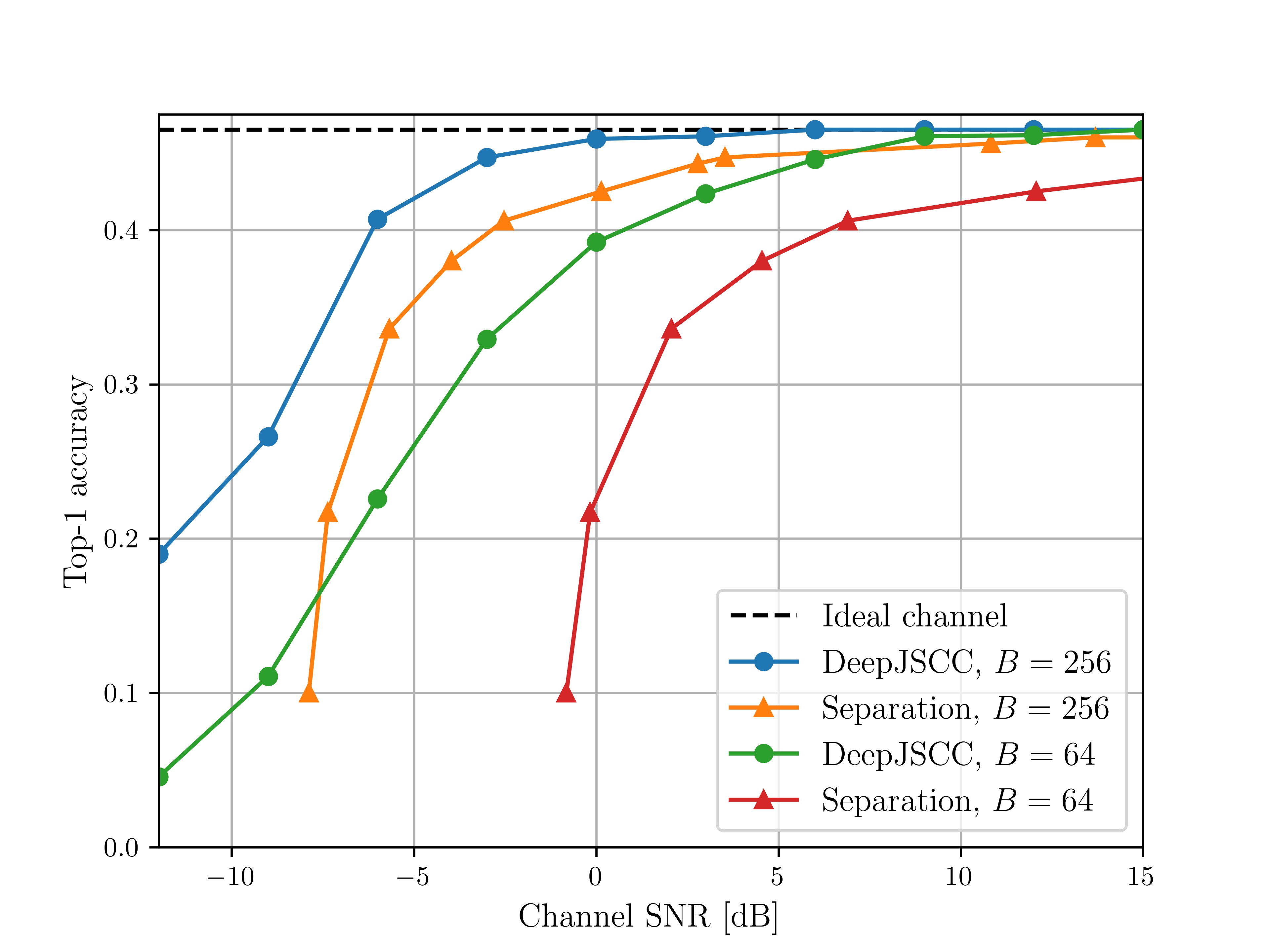}
\caption{\highlight{Top-1 accuracy performance of DeepJSCC scheme compared to the one achieved by a separation-based semantic communication approach assuming capacity-achieving channel coding.}}
\label{Fig:retrieval}
\end{figure}

\subsection{DeepJSCC for Image Retrieval}

In many emerging machine-type communication applications, rather than reconstructing the input signal, the receiver is interested in certain specific features, which, in general terms, represent the \emph{semantics} of the source for a particular downstream task. For example, the receiver may want to identify objects or other statistical properties in an image. In general, when the goal of the receiver is to recover a function of the input signal (e.g., classification or regression tasks), optimal performance can be obtained by the transmitter first estimating this function value, and then communicating its estimate to the receiver. However, in many cases, the transmitter may not have all the available information to make a reliable local decision. In those cases, it should extract the most relevant features of the source signal for the downstream task, and deliver them over the wireless channel.

In \cite{jankowski2021wireless}, a remote image retrieval problem is considered, where the objective of the receiver is to identify the subject in an image observed at the transmitter from a gallery of images available to the receiver. Note that, this decision cannot be made by the transmitter as it does not have access to the image gallery. As such, the objective of the JSCC problem here is classification, rather than reconstruction. The results are presented in Fig.~\ref{Fig:retrieval}, where the DeepJSCC approach is compared with a separation based scheme that combines retrieval oriented compression scheme with capacity achieving channel coding, where $B$ represents the channel bandwidth. We observe that the gap between the two is particularly large in the low-bandwidth low-SNR regime, even though the capacity-achieving channel coding assumption is very loose in this case. These results illustrate the significant benefits of DeepJSCC for low-power and ultra low latency applications.

\subsection{DeepJSCC for Effective/Pragmatic Communications}

In \cite{tungEffectiveCommunicationsJoint2021}, a new class of communication problems, called \textit{effective/pragmatic communications}, that generalize the JSCC/ semantic communication problem, is introduced.  An example is provided by considering an agent interacting with its environment through actions it takes. The agent does not observe the environment state, and relies on a remote controller for its actions. The controller can see the state perfectly, but has a noisy communication channel with limited bandwidth to communicate with the agent to help it take the right actions. In the framework proposed in \cite{tungEffectiveCommunicationsJoint2021}, the communication channel is considered as part of the environment dynamics, and the transmitted messages are treated as the actions taken by the controller. As a result, we obtain a multi-agent system, where the agents learn not only to collaborate with each other but also to communicate to accomplish their goal more efficiently. 
The authors show, via examples, that the joint policy learned using the proposed framework is superior to optimizing the actions within the environment and communication actions separately. By optimizing the objective of the task directly, the learned codewords transmitted by the controller show adaptivity to the underlying task, e.g., actions that are likely to lead to similar changes in the agent's state are mapped to similar codewords.

\section{Conclusion and Future Challenges} \label{sec:conclusions}

While the superiority of JSCC over separation in the practical finite block length regimes has been known, designing truly joint coding schemes that directly map an input signal to a channel codeword has been a long standing open research challenge. As we have shown here through examples DNNs can be successfully employed to design DeepJSCC schemes that can outperform state-of-the-art separation-based alternatives. We further emphasize that the latter are results of decades-long intensive research efforts, while DeepJSCC networks can be obtained by several hours of training. 

Another important benefit of the DeepJSCC paradigm is its computational efficiency. Even relatively shallow DNNs are sufficient to achieve satisfactory end-to-end performance, requiring significantly lower computational complexity compared to state-of-the-art image and video compression standards concatenated with iterative channel decoding schemes. Moreover, since they rely completely on DNN architectures, DeepJSCC schemes are inherently parallelizable. One should contrast this with the state-of-the-art in learning-aided channel code design. Despite significant research efforts in recent years, we still cannot design long block length channel codes in a purely data-driven manner. While DNN-based approaches have recently achieved state-of-the-art in image and video compression, the gains are still modest. This indicates that the JSCC problem is an easier problem from an end-to-end learning perspective, and the learning-based design approach of DeepJSCC holds a significant potential for the implementation of such codes in practice. 

While we have provided potential solutions for the security of DeepJSCC, we believe that security still constitutes an important challenge in front of its adoption in practical systems, particularly for sensitive applications. More research will be needed in this direction to incorporate more advanced security mechanisms into the design of DeepJSCC architectures. 

DeepJSCC is a powerful paradigm that can be applied to any semantic communication problem involving diverse source and channel statistics and any downstream task. We have mentioned wireless image and video delivery, CSI feedback, image retrieval, and multi-agent cooperation to demonstrate the flexibility of the DeepJSCC paradigm. On the other hand, an important challenge for DeepJSCC is to develop universal encoder/decoder architectures that can be used for multi-modal data sources, so that we do not need to train and store different network parameters for different source and channel combinations.  We also expect to see diverse applications of DeepJSCC for more challenging channels, such as optical, visible light, underwater, or satellite communication channels. In some of these scenarios, purely data driven approaches may need to be employed due to the lack of accurate channel models for training. 

Another potential application of the DeepJSCC paradigm is source delivery over multi-user networks. It is known that Shannon's Separation Theorem breaks down in most multi-user scenarios, even in the asymptotic infinite block length regime. This will require developing more advanced DeepJSCC techniques that can incorporate i) correlations among multiple source signals, ii) resource allocation techniques among nodes, and iii) interference management and cancellation techniques. 

Although favorable properties from superior performance to robustness against channel variations make DeepJSCC a promising technology in future communication systems, there are still many issues to be addressed before its adoption in practical systems. These include the peak-to-average power ratio (PAPR) problem when combining DeepJSCC with orthogonal frequency division multiplexing (OFDM) and the efficient training problem when DeepJSCC encoder and decoder adopt deeper and more complex networks. Solving these challenges will make DeepJSCC a key enabler of semantic communications in practical future wireless systems.  

\bibliographystyle{IEEEtran}
\bibliography{ref/ref.bib}
\vspace{-.5in}
\begin{IEEEbiographynophoto}
{Jialong Xu} is currently pursuing the Ph.D. degree with Beijing Jiaotong University. His research interests include wireless coding, semantic communications and information theory.
\end{IEEEbiographynophoto}
\vspace{-.52in}
\begin{IEEEbiographynophoto}
{Tze-Yang Tung} received  his PhD degree from Imperial College London. His current research is in joint source-channel coding and compression with emphasis on utilizing deep learning techniques to design novel algorithms.
\end{IEEEbiographynophoto}
\vspace{-.55in}
\begin{IEEEbiographynophoto}
{Bo Ai} is a Professor at Beijing Jiaotong University, China. His interests include OFDM, high-power amplifier linearization, radio propagation and channel modeling, and mobile communications for railway systems.
\end{IEEEbiographynophoto}
\vspace{-.55in}
\begin{IEEEbiographynophoto}
{Wei Chen} is a Professor with Beijing Jiaotong University, China. His current research interests include intelligent wireless communication systems and multimedia processing.
\end{IEEEbiographynophoto}
\vspace{-.5in}
\begin{IEEEbiographynophoto}
{Yuxuan Sun}is an Associate Professor with the School of Electronic and Information Engineering, Beijing Jiaotong University, China. Her research interests include edge computing, edge intelligence, and vehicular networks.
\end{IEEEbiographynophoto}
\vspace{-.5in}
\begin{IEEEbiographynophoto}
{Deniz G{\"u}nd{\"u}z} is a professor in the Electrical and Electronic Engineering Department, and leads the Information Processing and Communications Laboratory (IPC-Lab) at Imperial College London.
\end{IEEEbiographynophoto}
\end{document}